\documentclass[linenumbers]{aastex63}

\received{\today}
\revised{\today}
\accepted{\today}
\submitjournal{ApJ Letters}

\shorttitle{The Gravitational Redshift Observed in NGC 4258}
\shortauthors{Nucamendi et al.}

\begin{document}


\title{Towards the Gravitational Redshift Detection in NGC 4258 \\
and the Estimation of its Black Hole Mass-to-Distance Ratio}

\correspondingauthor{Ulises Nucamendi}
 \email{unucamendi@gmail.com}

\author[0000-0002-8995-7356]{Ulises Nucamendi}
\affiliation{Instituto de F{\'\i}sica y Matem{\'a}ticas, Universidad Michoacana de San Nicol\'as de Hidalgo,\\
Edificio C--3, Ciudad Universitaria, CP 58040, Morelia, Michoac{\'a}n, M{\'e}xico.}

\author[0000-0003-4918-2231]{Alfredo Herrera--Aguilar} 
\affiliation{Instituto de F{\'\i}sica, Benem{\'e}rita Universidad Aut{\'o}noma de Puebla,\\ 
Apartado Postal J-48, CP 72570, Puebla, Puebla, M{\'e}xico.}

\author[0000-0002-6170-8542]{Ra{\'u}l Lizardo--Castro} 
\affiliation{Instituto de Ciencias Nucleares, Universidad Nacional Aut{\'o}noma de M{\'e}xico,\\
Apartado Postal 70543, CP 04510, Ciudad de M{\'e}xico, M{\'e}xico.}

\author[0000-0002-1381-7437]{Omar L{\'o}pez--Cruz} 
\affiliation{Instituto Nacional de Astrof{\'\i}sica, {\'O}ptica y Electr{\'o}nica, (INAOE)\\
Luis E. Erro No. 1, Tonanzintla, CP 72840,  Puebla, M{\'e}xico.}











\begin{abstract}

We construct from first principles a general relativistic approach to study Schwarzschild black hole (BH) rotation curves and estimate the mass-to-distance ratio of the Active Galactic Nucleus (AGN) of NGC 4258 in terms of astrophysical observable quantities. The presented method allows one to clearly distinguish and quantify the general and special relativistic contributions to the total redshift expression.
The total relativistic redshift/blueshift comprises three components: the gravitational redshift due to the spacetime curvature generated by the mass of the BH in its vicinity, the kinematic shift, originated by the photons' local Doppler effect, and the redshift due to a special relativistic boost that describes the motion of a galaxy from a distant observer. 
We apply our method to the largest data set of highly redshifted water megamaser measurements on the accretion disk of the NGC 4258 active galaxy
and use this general relativistic method to estimate its BH mass-to-distance ratio: $M/D=(0.5326\pm 0.00022)\times 10^7 M_\odot{\rm Mpc^{-1}}$.


\end{abstract}



\section{Introduction} \label{sec:intro}

We have recently obtained strong observational evidences of the existence of BHs through both the gravitational wave detection  
\citep[e.g.,][]{2016PhRvL.116f1102A,2016PhRvL.116x1103A,2017ApJ...848L..12A} and the observation of the M87 BH shadow 
\citep[e.g.,][]{2019ApJ...875L...1E,2019ApJ...875L...4E,2019ApJ...875L...6E}. Besides, the observation of several star orbits of very short period around the center of the Milky Way envisages the presence of a supermassive black hole (SMBH) 
\citep[e.g.,][]{2003Sci...300.1898B,2005Natur.438...62S,2012RAA....12..995M}. Moreover, the GRAVITY collaboration has reported the observation of the Schwarzschild precession in the S2 star orbit at the center of the Milky Way by making use of a modified parameterized post-Newtonian formulation of General Relativity (GR) \cite{2020A&A...636L...5}. 
These new observations have motivated the advancing of relativistic methods to estimate  BH's mass and spin, considering GR or modified gravity theories \citep[See][and references therein]{2015PhRvD..92d5024H,2018PhRvD..97l4049S,2018arXiv180510512A}.

In many astrophysical studies that model the dynamics of stars, accretion disks, gas particles and other bodies orbiting a BH, the estimation the BH mass has been done by roughly identifying photons' redshift with the rotational velocities of the revolving objects $z=v_{rot}/c$. While this Newtonian approximation is fair when the orbiting bodies are far enough from the source, it breaks as soon as we approach the vicinity of the BH event horizon and the GR effects encoded in the photons' redshift start to manifest and become important. In other words, aside from the Doppler effect, photons' redshift also contains information about the spacetime curvature generated by the BH itself. Thus, in order to provide an appropriate account for the physics involved in such dynamical systems, it is convenient to fully express and physically interpret the redshift as a general relativistic invariant, especially in the vicinity of a BH event horizon.

In this Letter we present a new formalism for analysing very long baseline interferometry (VLBI) observations of megamasers \citep[see, e.g.,][for a general overview and the megamaser excitation mechanisms]{1980PhLB...93..107R,2005ARA&A..43..625L}  on the accretion disk of some nearby galaxies. We have considered the galaxy NGC 4258, which has been observed extensively for over three decades. The water maser spectrum of this galaxy displays almost symmetric bands of features about its systemic redshift, which is equivalent to a systemic velocity of about $500\;{\rm km\, s^{-1}}$: the highly redshifted features with velocities near $1500\;{\rm km\, s^{-1}}$ and the highly blueshifted features with velocities near $-500\;{\rm km\, s^{-1}}$ \citep[see review by ][and references therein]{2008ASPC..395...87M}. 
Redshifted water maser emission  was first detected in AGNs by \citet{1984Natur.310..298C}, and later from highly redshifted masers in NGC 4258 by  \citet{1993Natur.361...45N}. 
Further, by taking into account these highly redshifted features as well as apparent accelerations of the redshifted systemic masers, high precision VLBI studies provided direct evidence that the source of this spectral structure was a rapidly rotating disk in Keplerian motion where the systemic masers were confined to a narrow annulus with a fractional width of less than $0.2$, implying early estimations of the binding central mass of about $10^7$ solar masses \citep{1994ApJ...432L..35W} and $2.1\times 10^7$ solar masses  \citep{1995A&A...304...21G}. Thereafter, VLBA\footnote{VLBA is a highly precise astrometric machine with a resolution of 200 $\mu$as at the water maser frequency of 22 GHz which  possesses a very high correlator that renders a very fine spectral resolution, better than 1 km s$^{-1}$, allowing for the resolution of all of the NGC 4258 maser spectral features and the complete coverage of its spectrum.} observations demonstrated the full distribution of the masers in the rotating disk, where highly redshifted masers extend from 0.17 to 0.28 pc on either side of the systemic masers, and traced their rotation curve, revealing the disk's Keplerian motion to better than 1 \%. 
\citet{1995Natur.375..286S} performed a preliminary estimation of the gravitational redshift of the order of $2-4\;{\rm km\, s^{-1}}$ for the inner and outer edges of the disk, respectively, while several later attempts to model the disk in this system have included relativistic effects in a perturbative way (\citet{2005ApJ...629..719H,2013ApJ...775...13H,2019ApJ...886L..27R}).
These VLBA observations also indicated the presence of a central compact object with a mass of $3.6 \times 10^7\,M_\odot$, providing  evidence that NGC 4258 hosts a SMBH \citep{1995Natur.373..127M,2005ApJ...629..719H}.

Further, \citet{2005ApJ...629..719H} reported  a $2\sigma$ maximum deviation from Keplerian motion around the BH in the projected rotation curve of the NGC 4258 disk, an effect that amounts to a flattening of about $9\;{\rm km\, s^{-1}}$ in terms of velocities in the high-end of the velocity curve traced by the highly redshifted megamasers; these authors explained that such shift could be accounted for considering pure Keplerian motion in a model with an inclination-warped disk in which relativistic effects are not unambiguously detected.  VLBI observations  have been  used to determine geometric distance to NGC 4258 
\citep{2007ApJ...659.1040A,2005ApJ...629..719H,2008ApJ...672..800H,2013ApJ...775...13H}. The most recent results by  \citet{2019ApJ...886L..27R} reported a distance to NGC 4258 $D=7.576 \pm 0.082\: {\rm (stat.) \pm 0.076 (sys.)\; Mpc}$ and the Hubble constant equal to $H_0 = 72.0 \pm 1.9\;{\rm km\, s^{-1} \,Mpc^{-1}}$.

This paper is organized as follows: in \S1 we provide a general introduction, stressing the observations and previous studies of NGC 4258, in  \S2  we explain the general relativistic model employed in this paper, in \S3 we present the observations, \S4 presents the modeling, while in  \S5 and \S6 we present  discussions and conclusions, respectively. 
We finally provide a detailed mathematical derivation of the redshift associated with a special relativistic boost that encodes the receding motion of the NGC 4258 galaxy from us as well as the errors of the total shifts in the Appendix A.

\section{A General Relativistic  Method for Estimating Black Hole Mass-to-Distance Ratio}
 We present a simple general relativistic approach that allows us to estimate the BH mass in terms of observable  quantities:
\begin{itemize}
\item [ a)] The redshift of photons that are emitted by stars, galactic gas, water masers or other bodies revolving around the BH and travel along null geodesics towards a distant observer located on Earth, and 
\item [b)] The radius of the star/maser orbits which are assumed to be stable, circular and to lie on the equatorial plane.

\end{itemize}

Let's consider a Schwarzschild BH  whose known exact solution has mass $M$ and event horizon radius $r_h=2GM/c^2\equiv 2m$ (in natural units):
\begin{equation}
ds^2\!=\!\frac{dr^2}{f} + r^2\!\left(d\theta^2\!+\!\sin\!^2\theta d\varphi^2\right)\!-\!f dt^2, \qquad f\!=\!1- \frac{2m}{r}.
\label{Schmetric}
\end{equation}
In particular, this metric possesses two commuting {\em Killing vectors} responsible for the  conservation of the energy $E$ and the component of angular momentum parallel to the symmetry axis $L$ \citep[e.g.,][]{1972ApJ...178..347B,2001PhRvD..63l5016N,2004PhRvL..92e1101L}.

Objects which emit or detect photons geodesically move around BH with 4-velocities $U^{\mu} = (U^{t}, U^{r}, U^{\theta}, U^{\varphi})$,  normalized to unity: $U^\mu U_\mu=-1$. This condition renders a relation that resembles the Energy Conservation Law for a non-relativistic particle with energy $E^2/2$ and unit mass that moves in a potential 
\begin{equation}
V_{eff}=\frac{f}{2}\left[1+\frac{L^2}{r^2\sin^2\theta}+r^2(U^{\theta})^2\right].
\end{equation} 
Considering  equatorial and circular orbits ($U^{\theta}\!=\!0\!=\!U^r$), this effective potential must obey the following conditions: $V_{eff}=0=V^{\prime}_{eff}$, with the supplementary stability restriction $V^{\prime\prime}_{eff}<0$ \citep{1972ApJ...178..347B}, which in the language of  radial coordinate reads $r\!>\!6m$; here primes denote derivatives with respect to $r$.

Similarly, the photons emitted by stars and/or galactic gas possess 4-momentum $k^{\mu} = (k^{t}, k^{r}, k^{\theta}, k^{\varphi})$ and travel towards us along null geodesics ($k^{\mu}k_{\mu} =0$) preserving both energy and axial angular momentum: $E_{\gamma}$ and $L_{\gamma}$.

The Schwarzschild redshift and blueshift (respectively corresponding to receding and approaching emitting sources and labeled by the subscripts $_{{\rm Schw}_{1,2}}$) that these photons experience in their path from the equatorially and circularly orbiting bodies towards a static distant observer read
\citep{2015PhRvD..92d5024H} 
\begin{equation}
1+Z_{{\rm Schw}_{1,2}}\equiv\frac{\omega_e}{\omega_d}\!=\!\frac{(k_{\mu}U^{\mu})\!\mid_{e}}{(k_{\mu}U^{\mu})\!\mid_{d}}=
\frac{(U^{t} - b_{_\mp} \,U^{\varphi})\mid_{e}}{(U^{t} - b_{_\mp} \,U^{\varphi})\mid_{d}},
\label{z}
\end{equation} 
where latin subscripts label the emitter($_{e}$) and the detector($_{d}$)  and $b\equiv L_{\gamma}/E_{\gamma}$ is the light bending  (deflection) parameter \footnote{Also called the  {\em impact parameter} in previous works.}, a quantity that is defined by the BH gravitational field, remains constant along the whole photons' path $b_{e}\!=\!b_{d}$, and is given by 
\begin{equation} 
\label{b} 
b_{_\mp} = \mp\sqrt{-\frac{g_{\varphi\varphi}}{g_{tt}}} = \mp\sqrt{\frac{r_e^3}{r_e-2m}}.
\end{equation}
In this maximized expression we have considered that photons are emitted at the trajectory point where $k^r=0$ \citep{2001PhRvD..63l5016N,2004PhRvL..92e1101L};
here the $(_{_\mp})$ subscripts indicate that there are two different values of the light deflection parameter that account for the approaching/receding movement of the stars/gas with respect to the distant observer either side of the line of sight (LOS) (The LOS links the BH with the distant observer, with the Local Standard of Rest, LSR, in our case), yielding the corresponding Schwarzschild redshifts $Z_{{\rm Schw}_1}$ and $Z_{{\rm Schw}_2}$ different from each other, both in sign and magnitude.

For equatorial and circular orbits, we have the following expression: 
\begin{eqnarray}
U^{t}(r, \pi/2) 
=\sqrt{\frac{r}{r-3m}},   \qquad 
U^{\varphi}(r, \pi/2) 
=\pm\frac{\sqrt{m}}{r \sqrt{r - 3m}}, 
\label{Ut}
\end{eqnarray}
where the $\pm$ signs reveal the election of the angular velocity direction.
When the orbit of the detector stands far away from the BH we can consider it as static and the 4-velocity simplifies: 
\begin{equation} 
\label{Ustatic} 
U^\mu|_d=(1,0,0,0), \qquad \textrm{when}  \qquad  r_d\to\infty.
\end{equation}

Thus, from (\ref{z})-(\ref{Ut}) the Schwarzschild redshifts read
\begin{equation} 
\label{Z1,2} 
1 + Z_{{\rm Schw}_{1,2}} = \sqrt{\frac{r_e}{r_e-3m}} \pm\sqrt{\frac{m r_e}{\left(r_e\!-\!2m\right)\left(r_e\!-\!3m\right)}}\,,
\end{equation}
here the $\pm$ signs correspond to the clockwise and counter-clockwise motion of the probe particles with respect to a distant observer, respectively. In general, $|Z_{{\rm Schw}_1}| \neq |Z_{{\rm Schw}_2}|$ mainly because of the gravitational redshift due to the space-time curvature generated by the presence of the BH mass; we should note that in the Newtonian picture this spacetime curvature is interpreted as a force or acceleration acting on massive particles, but within general relativity these particles move along geodesics and can be both massless or massive. 
These quantities express the redshifts and blueshifts experienced by photons which travel along null geodesics and are emitted by circularly orbiting bodies in the equatorial plane around a Schwarzschild BH towards a static observer located far away from it.

Let us define the {\it gravitational redshift} as the first term of Eq. (\ref{Z1,2}) 
\begin{equation}
1+Z_{{\rm grav}}
=U^{t}_{e}=
\sqrt{\frac{r_e}{r_e-3m}},
\label{Zc}
\end{equation} 
and the kinematic redshifts $Z_{{\rm kin}_\pm}$ either side of the BH source as the second item of Eq. (\ref{Z1,2}) (it can be defined as well by subtracting from the Schwarzschild redshift $Z_{{\rm Schw}_{1,2}}$ the central redshift $Z_c\equiv Z_{{\rm grav}}$ experienced by photons emitted by a particle located at the LOS, where $b=0$):
\begin{eqnarray} 
\label{Z1} 
Z_{{\rm kin}_\pm}\!
\equiv Z_{{\rm Schw}_{1,2}}\!-\!Z_{{\rm grav}}=
\pm\, U^\varphi_e\, |b_{_\mp}|\!=\!
\pm\sqrt{\frac{m r_e}{\left(r_e\!-\!2m\right)\left(r_e\!-\!3m\right)}}\,.
\end{eqnarray}
In Eqs. (\ref{Z1,2})-(\ref{Zc}) we have taken into account Eq. (\ref{Ustatic}).

We further compose the Schwarzschild shift (\ref{Z1,2}) with the redshift describing the motion of a galaxy from a distant observer, $Z_{{\rm boost}}$, which is associated with a special relativistic boost  \citep[][]{1984ucp..book.....W}
\begin{eqnarray}
  \label{Zboost1}
  1 + Z_{\rm boost}= \gamma (1 + \beta), \qquad  \gamma \equiv \left(1-\beta^2\right)^{-1/2}, \qquad \beta \equiv \frac{v_{0}}{c}, \qquad v_0 = Z_0 c,
\end{eqnarray}
where $v_0$ is the systemic velocity of the galaxy with respect to a distant observer and $Z_0$ is the systemic redshift, 
rendering the total redshift \citep[e.g.,][we provide a  derivation of this result in Appendix \ref{sec:appendix}]{2014MNRAS.442.1117D}:
\begin{equation}
 Z_{\rm tot_{1,2}} = (1 + Z_{\rm Schw_{1,2}}) (1 + Z_{\rm boost}) -1 = 
 \left( 1+Z_{{\rm kin}_\pm} + Z_{{\rm grav}} \right) \gamma (1 + \beta) -1.
\label{Zt}
\end{equation}
By performing a series expansion of the total redshift under the conditions $Z_0\ll 1$ ($v_0\ll c$) and $m\ll r_e$ we find 
\begin{eqnarray}
Z_{\rm tot_{1,2}}&\approx&
\left( 1+Z_{{\rm kin}_\pm} + Z_{{\rm grav}} \right) \left(1 + Z_0-\frac{Z_0^2}{2}\right) -1 
\approx
Z_{{\rm kin}_\pm} + Z_0 + Z_{{\rm grav}} + Z_0 Z_{{\rm kin}_\pm} + Z_0 Z_{\rm grav} + \frac{Z_0^2}{2} 
\nonumber\\
&\approx&
\pm\sqrt{\frac{m}{r_e}} + Z_0 +\frac{3}{2}\frac{m}{r_e} \pm\sqrt{\frac{m}{r_e}}Z_0 + \frac{3}{2}\frac{m}{r_e}Z_0 \pm\frac{5}{2}\left(\frac{m}{r_e}\right)^{3/2} + \frac{Z_0^2}{2} ,
\label{series} 
\end{eqnarray}
an expression that reveals that the Keplerian Doppler component of the redshift is the dominant one (it also coincides with the leading term in the expansion given by Equation (\ref{Z1}) of the kinematic redshift), while the next-to-leading order item is the redshift corresponding to the special relativistic boost that encodes the motion of the galaxy with respect to a distant observer, the third term is the gravitational redshift generated by the spacetime curvature due to the presence of the BH mass, the fourth and fifth items correspond to special relativistic corrections to the kinematic and gravitational redshifts, respectively.
It is straightforward to see that this expression contains all the relativistic corrections computed in Eq. (4) of \citet{2005ApJ...629..719H} for the NGC 4258 dynamical system. In fact, for this AGN all relativistic corrections of order greater than $(m/r_e)^1$ in  Eq. (\ref{series}) impart adjustments to the predicted maser motions at a negligible level ($\ll 1$ km s$^{-1}$, in terms of velocities), therefore rendering them unobservable since the individual measurement uncertainties are on the order of $1$ km s$^{-1}$; previous efforts reported in the literature included all effects up to order $(m/r_e)^1$.

Thus, the overall meaningful  relativistic redshift effect produced by the sum of the gravitational redshift $Z_{{\rm grav}}$ and the special relativistic correction to the kinematic redshift $Z_{{\rm sprel}}\equiv Z_0 Z_{{\rm kin}_\pm}$ (hereinafter called kinematic boosted redshift) reads
\begin{equation} 
\label{Zrel} 
Z_{{\rm rel}}\equiv Z_{{\rm grav}} + Z_{{\rm sprel}}.
\end{equation}

\section{Observations of Megamasers on  the NGC 4258 Accretion Disk }

In this work we consider VLBI  observations of the   megamaser system in the accretion disk in  NGC 4258, which consists of an AGN with a SMBH surrounded by a rotating accretion disk of subparsec diameter that contains several water masers in nearly Keplerian motion.  We use data  reported by \citet{2007ApJ...659.1040A} that account for the redshifts/blueshifts of photons emitted by water masers at the points of maximal emission with certain orbital radii. The adopted center is at  $\alpha_{2000}={\rm 12^h\,18^m}\,57\fs5046\, \pm 0\fs0003$, $\delta_{2000}= 47\arcdeg\,18\arcmin\,14\farcs303\, \pm 0\farcs003$ \citep{2005ApJ...629..719H}.
\citet{2007ApJ...659.1040A} have reported  observations  taken from 1997 March 6 through 2000 August 12, using the following  facilities, the Very Long Baseline Array (VLBA)\footnote{The VLBA is a facility of the National Radio Astronomy Observatory, which is operated by Associated Universities, Inc., under cooperative agreement with the National Science Foundation.}, the Very Large Array (VLA), and the Effelsberg 100 m telescope (EFLS) of the Max-Planck-Institut f\"ur Radioastronomie. Twelve ``medium-sensitivity" epochs were carried with the  VLBA alone; while, six  ``high-sensitivity" epochs  were interleaved and involved the VLBA, augmented by the phased VLA and EFLS. Calibration and synthesis imaging were done using the standard package Astronomical Image Processing System (AIPS)\footnote{\url{http://www.aips.nrao.edu/index.shtml}}. 
The VLBA  provides an angular resolution of 200 micro arcsec ($\mu$as) and a spectral resolution of ${\rm 1\;km\,s^{-1}}$ at the water maser frequency of 22 GHz.

\section{A New Modeling of the NGC4258 Megamaser Dynamical  System }

The NGC 4258 dynamical system closely approximates the assumed conditions that led us to the redshift Equations (\ref{Z1,2})-(\ref{Zc}) since the water maser clouds on the accretion disk describe circular orbits around the BH, lying  almost in the equatorial plane as we see this system nearly edge-on from Earth \citep[see][for details]{2008ASPC..395...87M,2005ApJ...629..719H}. Other promising  megamasers systems on edge-on accretion disk have been detected  \citep[e.g.,][]{2008ASPC..395..103B, 2013ApJ...767..154R,2017ApJ...834...52G}.

We minimized the $\chi^2$ by a Bayesian statistical fit based on Markov-Chain Monte Carlo scheme applied to the rotation curves traced by the megamasers using the formalism developed above (\S2). We are applying our fits to  directly measured/observed general relativistic invariant quantities, the redshifts.

We fit for the following  three parameters: The BH mass-to-distance ratio $M/D$, the redshift of the galaxy as a whole (the systemic velocity) $Z_0$ and the offset $x_0$ of the center of mass position $r_0(x_0,y_0)$. The origin of coordinates $(x_0,y_0)$ in the Newtonian approach corresponds to the origin $(0,0)$ in the equatorial plane of our general relativistic approach; here we use the Newtonian origin, where the disk center is measured with respect to a reference systemic maser at $510\;{\rm km\,s^{-1}}$. Since due to the east-west orientation of the extremely thin disk, the highly redshifted fit is insensitive to $y_0$, therefore we fixed $y_0=0.556$ mas as estimated by \citet{1997ApJ...475L..17H}, based on spatial symmetry considerations of the highly redshifted megamasers. We have performed several statistical fits shifting the $y_0$ value by $\Delta y$ of one and two orders of magnitude greater than the $x_0$ reported errors and our results remained practically unchanged, rendering almost the same $M/D$ and $\chi^2_{\rm red}$ values.

Additionally, we shall assume  that not all masers lie along the midline, but are uniformly scattered about it by an independent of radius azimuthal angle $\delta\varphi=5^{\circ}.9$, analogously to \citet{2005ApJ...629..719H}, where $\varphi=\pi/2$ defines the midline, and that the disk inclination is parametrized  by the polar angle $\theta_0=82^{\circ}$ (where $\theta=\pi/2$ corresponds to an edge-on view from Earth). 

For highly redshifted/blueshifted (high-velocity) masers lying on the midline of the inclined disk, which is receding from us with a systemic redshift $Z_0$\footnote{Were this galaxy within the Hubble flow we would need to apply an extra composition of redshifts in order to account for the cosmological redshift.}, the general relativistic model (\ref{Zt}) for fitting the observational data is 
\begin{eqnarray}
\label{Z1,2i} 
\widetilde{Z}_{{\rm tot}_{1,2_i}}=
\left[\sqrt{\frac{r_{e_i}}{r_{e_i}-3m}}\left(1\pm\sqrt{\frac{m}{\left(r_{e_i}-2m\right)}}\sin\theta_0\right)\right]
\frac{1+Z_0}{\sqrt{1-Z_0^2}}-1,
\end{eqnarray}
where $r_{e_i}$ is the orbital radius of the $i$-th maser. We should stress that we work with the full general relativistic redshift instead of Newtonian velocities ($v_i$) derived from the optical definition of the Doppler shift (or kinematic redshift) $v_i/c=(\omega_e-\omega_d)/\omega_d$, where now $\omega_e$ labels the rest frequency emitted at the water transition (22.23508 GHz) and $\omega_d$ is the measured by the detector frequency of the $i$-th maser.

Following the idea of \citet{2005ApJ...629..719H}, the $\chi^2$ of the general relativistic model we fit reads
\begin{equation}
\label{chi2} 
\chi^2=\sum_i \frac{\Big[Z_i - \epsilon\sin\theta_0\left(1+Z_{\rm boost}\right) \, Z_{{\rm kin}_\pm} - Z_{\rm grav} - Z_{\rm boost} - Z_{\rm grav}Z_{\rm boost}\Big]^2}
{\sigma^2_{Z_i} + \beta^2 Z_{{\rm kin}_\pm}^2},
\end{equation}
where the summation is performed over all highly redshifted maser features $Z_i$ and $\sigma_{Z_i}$ are the combined uncertainties in the model and observed redshifts, given by
\begin{equation}
\label{sigma2zi} 
|\sigma_{Z_i}| \equiv \delta Z_{{\rm tot}_{1,2}}\,,  \end{equation}
where $\delta Z_{{\rm tot}_{1,2}}$ is defined by Eq. (\ref{errorTotal}) and $\frac{\delta r_e}{r_e}\approx \frac{\sigma_{x_i}}{x_i-x_0}$, $\sigma_{x_i}\approx \Theta_x/(2S/N)$ is the right ascension uncertainty of the $i$-th feature, $\Theta_x$ is the $x$-component of the synthesized beam, $S/N$ is the signal-to-noise ratio, and there is no $\sigma_{y_i}$ contribution to $\sigma_{Z_i}$ because of the east-west orientation of the disk and the earlier mentioned fit insensitivity to $y_0$; the parameters $\epsilon$ and $\beta$ account for the above mentioned scatter about the midline, and are related to $\delta\varphi$ as follows:
\begin{equation} 
\label{eb} 
\epsilon\approx 1-\frac{\delta\varphi^2}{2}+\frac{\delta\varphi^4}{24}, \qquad \beta^2\approx\frac{\delta\varphi^4}{4},
\end{equation}
where the first expansion corresponds to the cosine function of the azimuthal angle $\varphi$ and the second one to the induced uncertainties of the maser scatter under the assumption that $\varphi\ll 1$ and $\varphi\approx\delta\varphi$.

\begin{figure}
\begin{tabular}{ c c}
    \includegraphics[width=9cm]{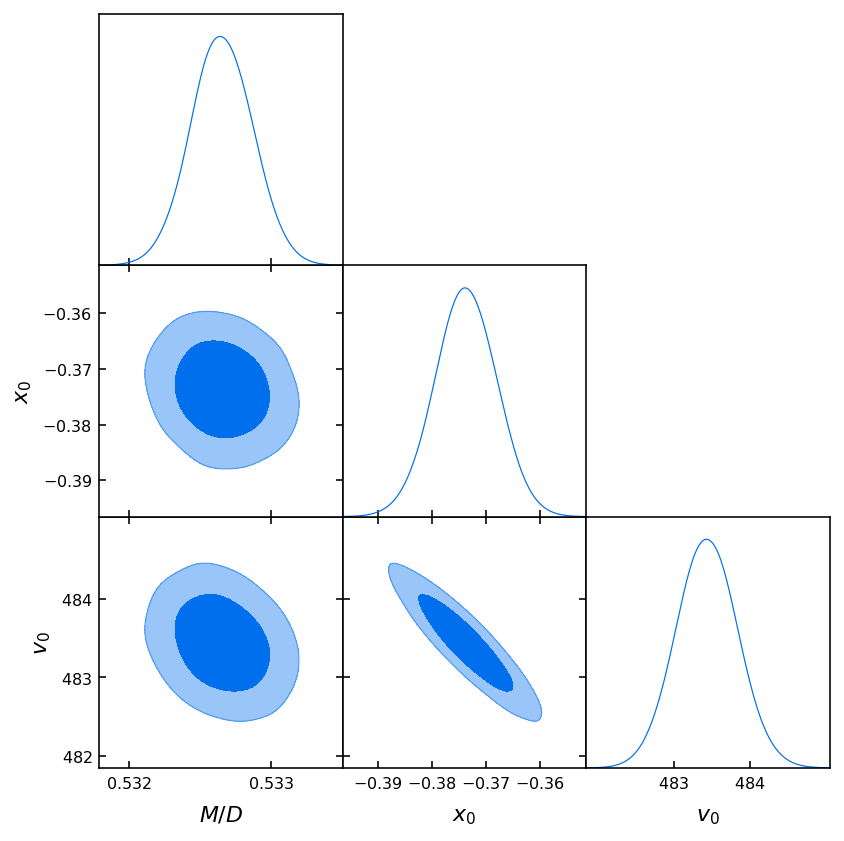} 
   & \includegraphics[width=9cm]{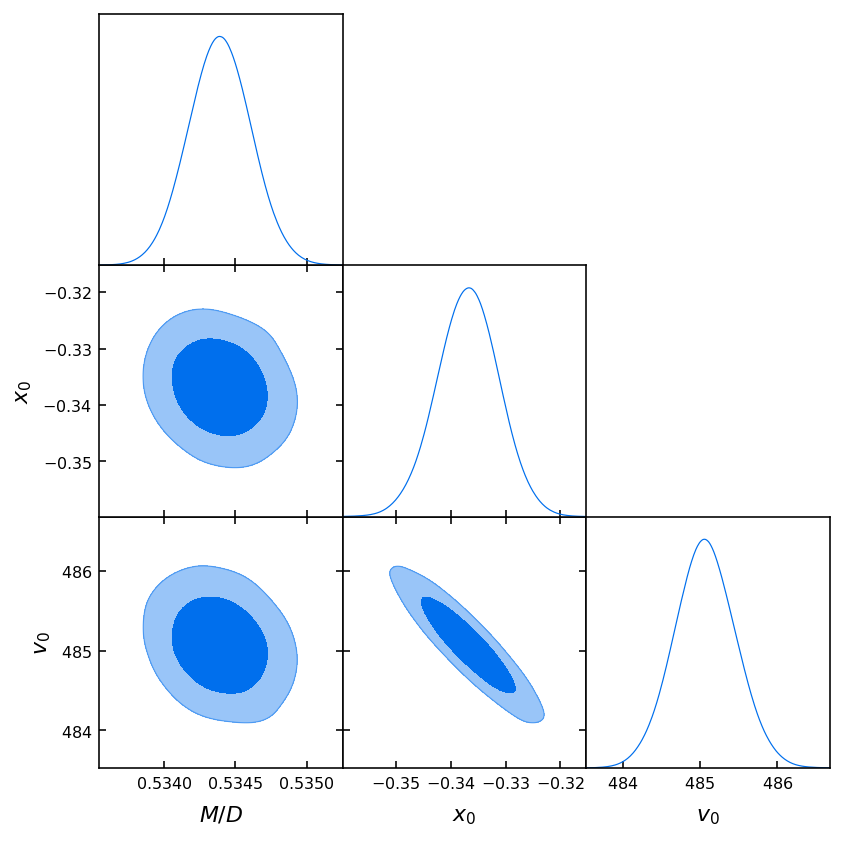}\\
   a) & b) 
   \end{tabular}
\caption{The posterior distribution of the general relativistic [Newtonian] Bayesian fit is displayed in panel a) [b)]. Here the BH mass-to-distance ratio $M/D$ is expressed in $\times 10^7 M_\sun{\rm Mpc^{-1}}$, while $x_0$ is expressed in mas; $Z_0$ is a dimensionless quantity. Contour levels correspond to $1\sigma$ and $2\sigma$ confidence regions.}\label{fig1}
\end{figure}

We have performed both a GR and a Newtonian Bayesian statistical fits for the aforementioned model parameters $M/D$, $Z_0$ $(v_0)$ and $x_0$ under the same assumptions and using the following Newtonian model
\begin{equation}
\label{ZN1,2i} 
v_{{\rm Newt}_{1,2_i}}=\pm\sqrt{\frac{m}{r_{e_i}}}\sin\theta_0+v_0, \qquad  \qquad v_0\equiv Z_0\,c.
\end{equation}
The results are presented in Table \ref{tab1} and Fig. \ref{fig1}. Both fits render very similar, but excluding each other best estimation values for the parameters with a slight preference for GR in terms of the $\chi^2_{\rm red}$, an effect due to the fact that the difference between both theories is slightly less than $1\%$.

\begin{deluxetable}{llllc}
\tablecolumns{5}
\tablewidth{0pt}
\tablecaption{General Relativistic and Newtonian Fits to NGC 4258 Megamaser System \label{tab1}}
\tablehead{
\colhead{Model} & \colhead{$M/D\; (\times 10^{7}\;M_{\sun}{\rm Mpc^{-1}}$)}& \colhead{$Z_0 (\times 10^{-6})$} &  \colhead{$x_0 \, ({\rm mas})$} & \colhead{$\chi^2_{\rm red}$}\\
&&\colhead{$v_0\; ({\rm km\, s^{-1}}$)}&&\\}
\startdata
General & $0.5326\pm 0.00022$ & $1612.6 \pm 1.3$ & $-0.374 \pm 0.006$ & 1.1988  \\
Relativistic & & $483.44 \pm 0.40$  & & \\
\hline
Newtonian & $0.5344 \pm 0.00022$ & $1618.0 \pm 1.3$ & $-0.337 \pm 0.006$& 1.2188  \\
& & $485.06 \pm 0.40 $          &                                      &    \\
\hline
\enddata
\tablecomments{The BH mass-to-distance ratio, the systemic redshift (velocity), the offset in the east-west direction of the center of mass BH position, and the reduced $\chi^2_{\rm red}$. We have used flat prior distributions defined in appropriate intervals.
}
\end{deluxetable}

\begin{figure}
\begin{center}
\includegraphics[width=12cm]{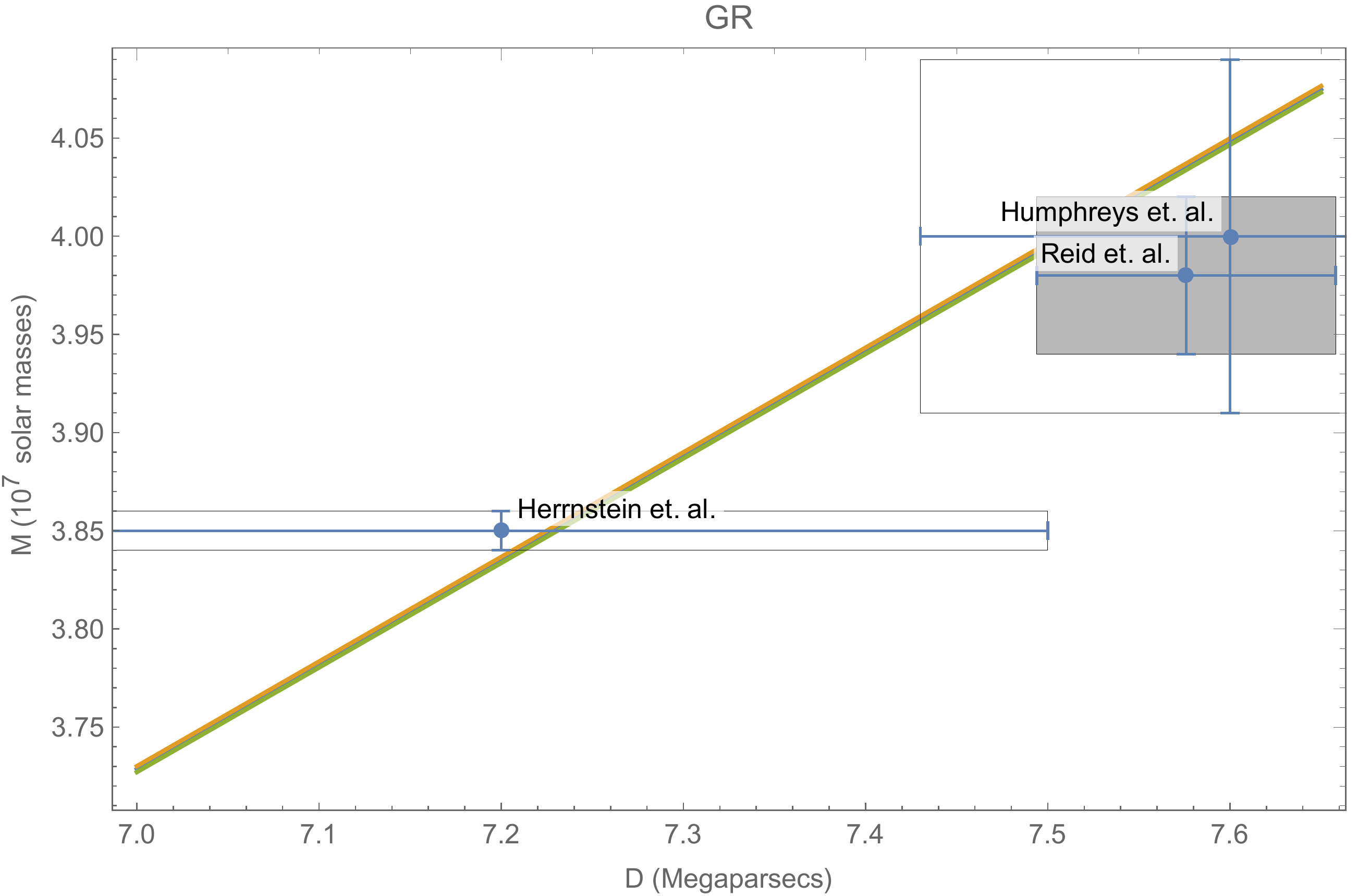}
\end{center}
\caption{The $M$ {\it vs.} $D$ plot shows the $1\sigma$ restriction on the mass-to-distance ratio general relativistic estimation. It is in agreement with the $1\sigma$ restriction on the more recent best estimation value for the distance to the source reported by \citet{2019ApJ...886L..27R} as the intersection of our restriction fringe with the shadowed region indicates.}
\label{md}
\end{figure}

We now substitute the GR best fit values of the BH mass-to-distance ratio $M/D=0.5326\times 10^7 M_\odot{\rm Mpc^{-1}}$ (see Fig. \ref{md}) and the systemic redshift $Z_0=1612.6\times 10^{-6} \ (v_0=483.44 \ \rm km\, s^{-1})$ as well as the radius of the closest to the BH redshifted maser, $r_{e_i}\sim 3.17 \ {\rm mas}$, into the expression of the relativistic redshift (\ref{Zrel}) and obtain a quantified value
\begin{equation}
Z_{{\rm rel}} =0.00002487 + 0.0000065 = 0.00003137, \qquad \qquad v_{\rm rel} = (7.45 + 1.96)\: {\rm km\, s^{-1}} = 9.41\: {\rm km\, s^{-1}},
\label{Zrelvalue} 
\end{equation}
that accounts for the maximum relativistic deviation from pure Keplerian motion in NGC 4258.

\begin{figure}
\begin{tabular}{ c c}
    \includegraphics[width=9cm]{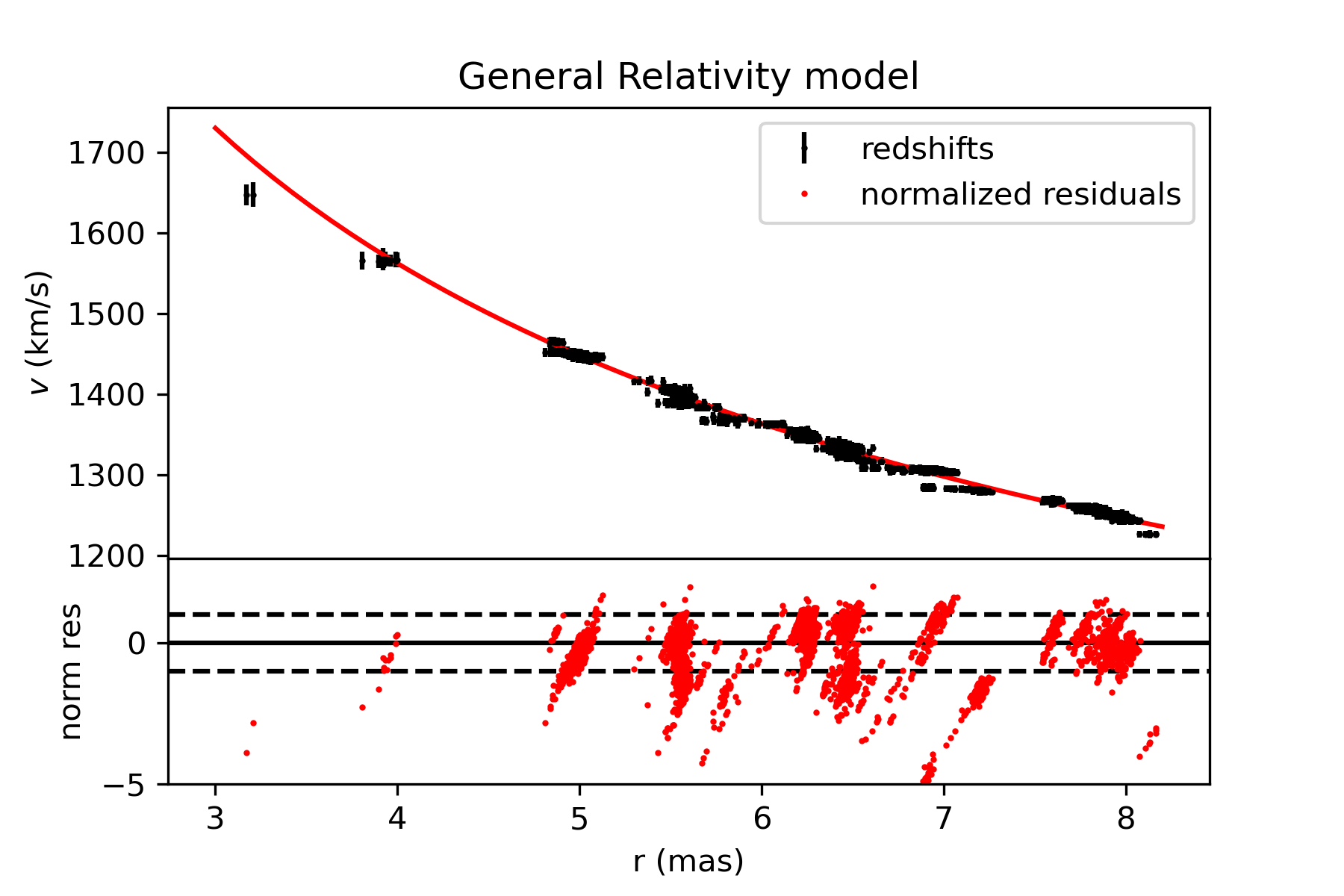} 
   & \includegraphics[width=9cm]{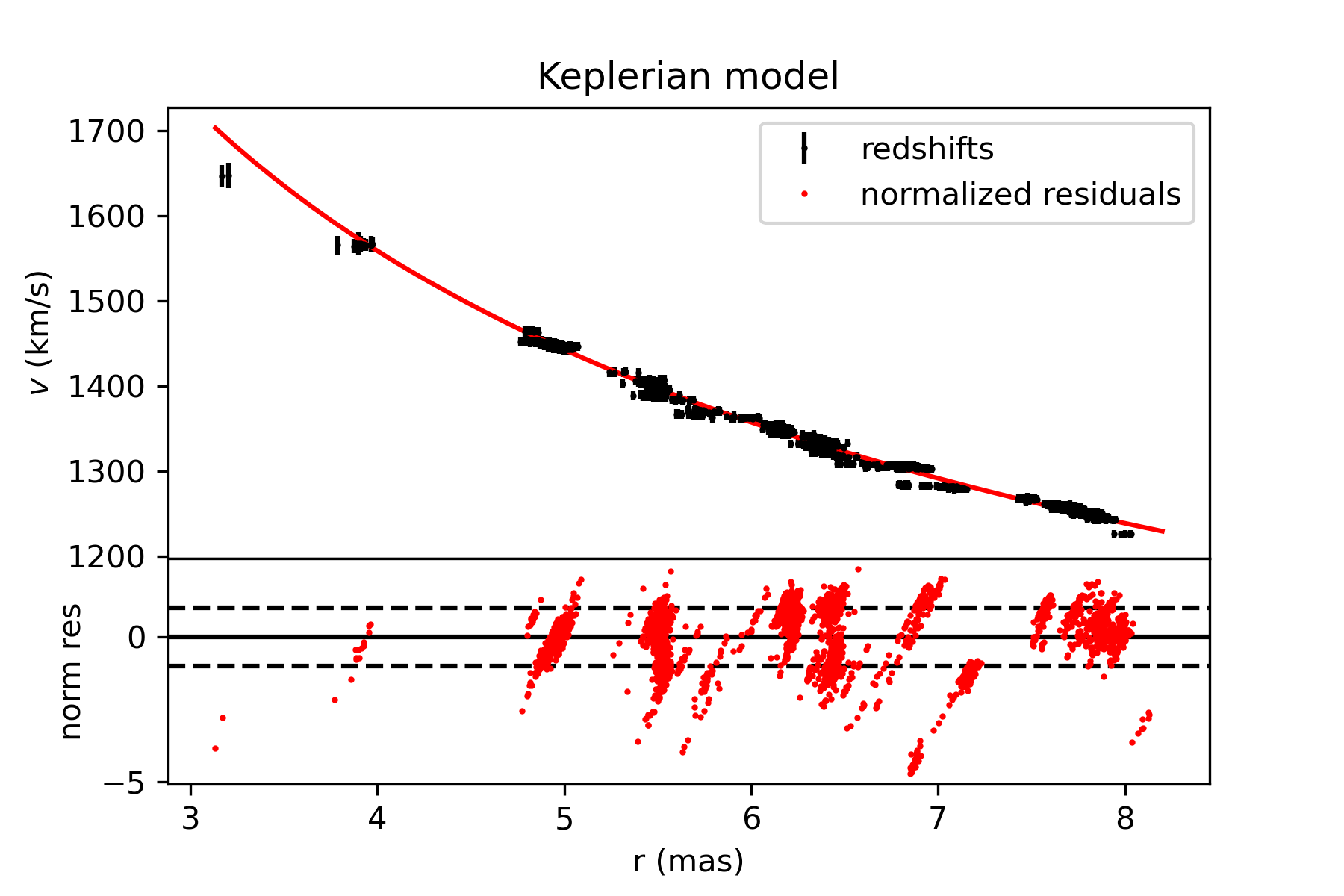}\\
   a) & b) \\
    \includegraphics[width=9cm]{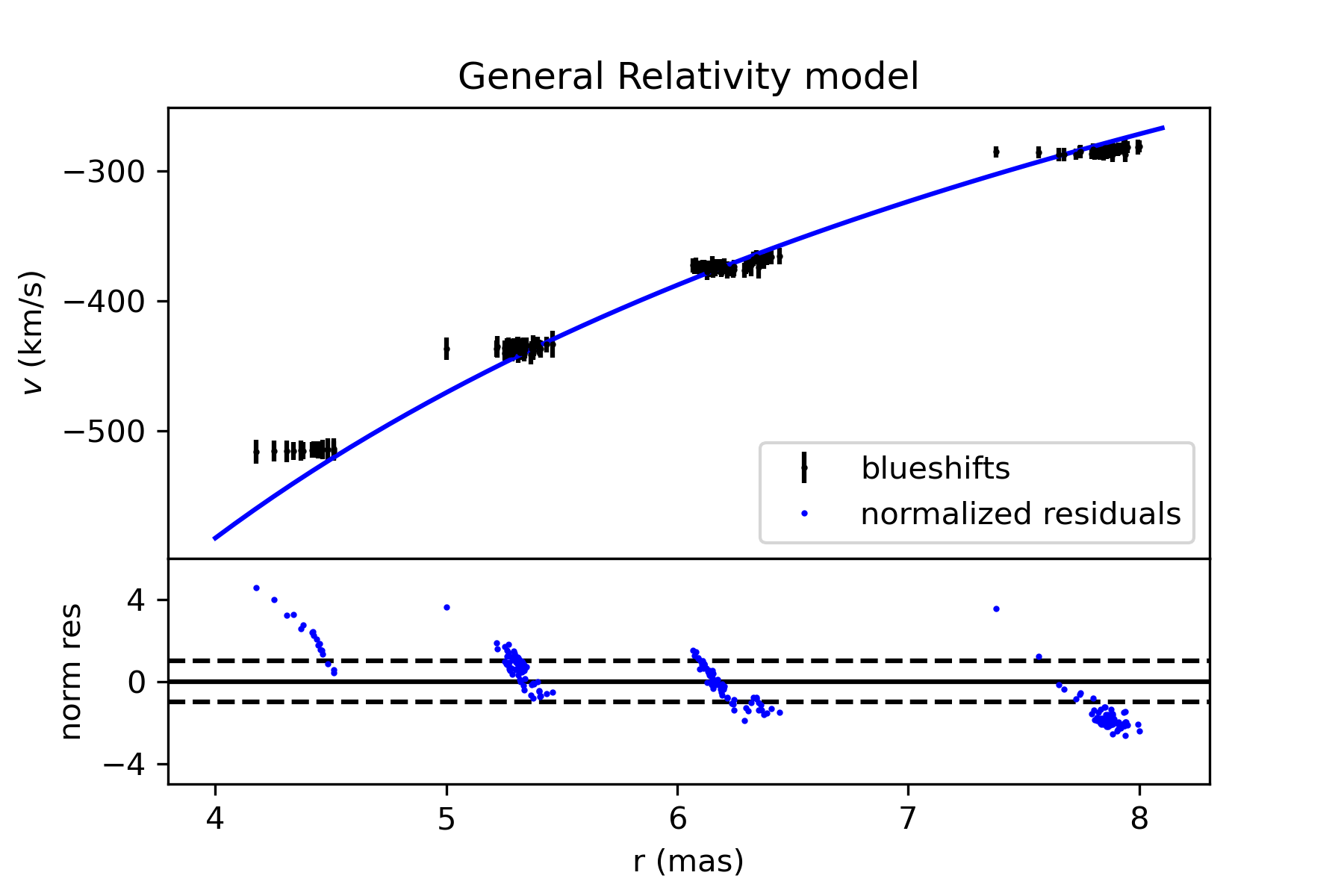}
   & \includegraphics[width=9cm]{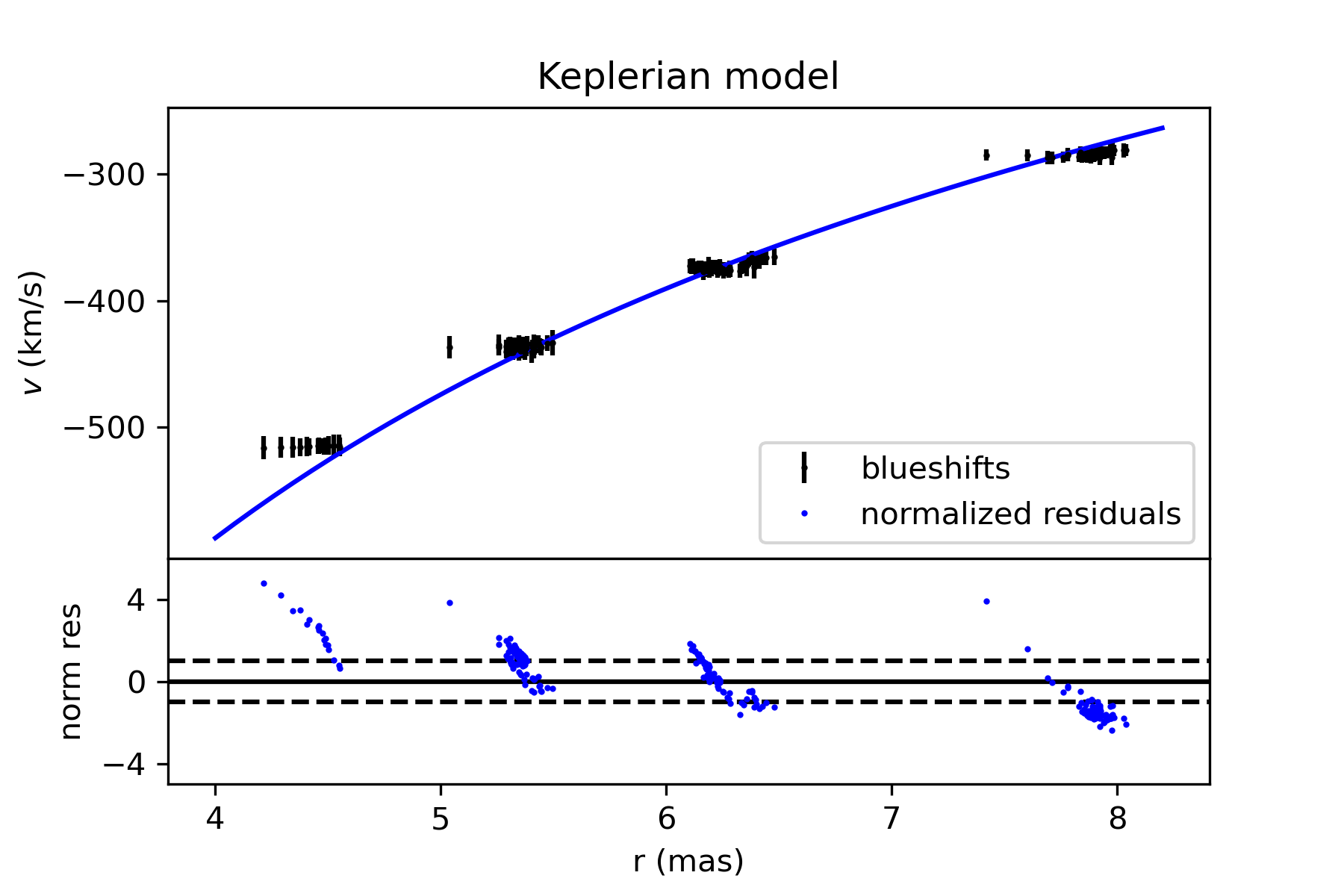}\\
   c) & d)
   \end{tabular}
\caption{Here we display both the general relativistic [panels a) and c)] and Newtonian [panels b) and d)] best fitting rotation curves of highly redshifted and blueshifted photons emitted by water masers in their disk motion around the galactic center during 18 epochs spanning 3 years reported by \citet{2007ApJ...659.1040A} The upper subpanels show radial positions and redshifts/blueshifts of maser features with their corresponding error bars. 
Normalized residuals are displayed in the lower subpanels.
}\label{fig3}
\end{figure}

In Fig. \ref{fig3} we plot both the general relativistic and the Newtonian best fitting projected rotation curves of the redshifted and blueshifted features, using the data of the 18 epochs spanning 3 years reported by \citet{2007ApJ...659.1040A}, together with the corresponding normalized residuals.


\section{Discussion}

\citet{2005ApJ...629..719H} presented  several alternative models that deviate from pure Keplerian motion (non-Keplerian rotation, a spherically symmetric central cluster of objects, a massive disk and a warped inclined disk) in order to account for the above mentioned $2\sigma$ maximum deviation. 

The asymmetry effect in the rotation curves of the highly red- and blueshifted masers was reported in \citet{Herrnstein-Thesis}, but was missing in the \citet{2005ApJ...629..719H} approach. In Fig. 3b these authors display $|v_{\rm los_i}-v_0|$ and show a single rotation curve for both redshifted and blueshifted maser’s photons, from which it is not possible to observe the aforementioned asymmetry that encodes the gravitational redshift effect. Our expression for the total redshift is asymmetric by definition since the contribution of the gravitational redshift is positive definite, in contrast to the kinematic shift which is positive for redshifted masers and negative for blueshifted ones; these kinematic shifts are equal in magnitude. 

\begin{figure}
\begin{center}
    \includegraphics[width=12cm]{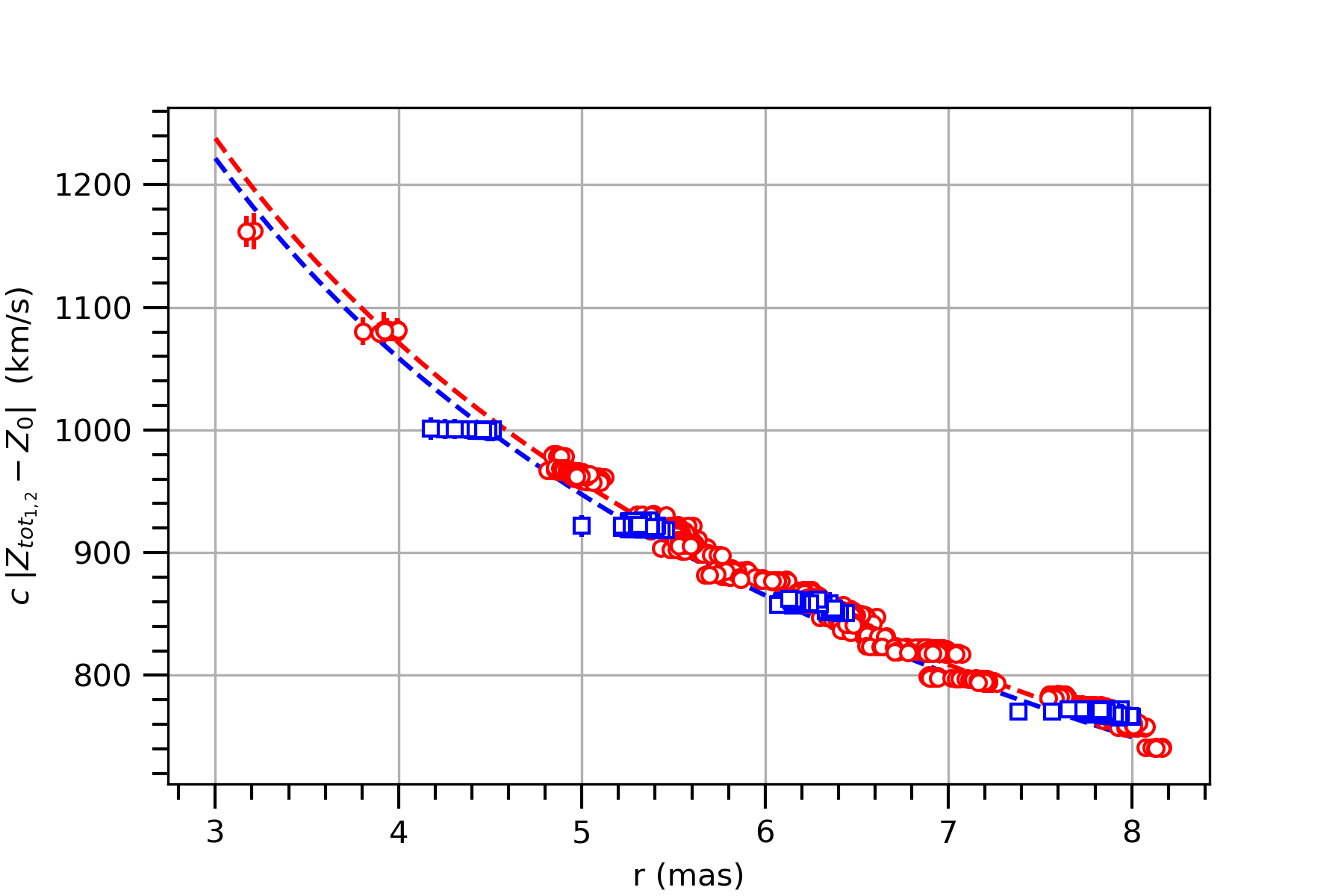} 
   \end{center}
\caption{Here we display the overlaying of the rotation curves of the highly red- and blueshifted masers where the special relativistic redshift has been approximated by the systemic one and has been subtracted from the total shift in order to make evident the asymmetric effect due to the gravitational redshift. The red circles denote redshifted features while the blue squares represent blueshifted ones with their respective error bars.
}
\label{fig4}
\end{figure}

In Fig. \ref{fig4} the effect of the asymmetry is evident as the separation between the red- and blueshifted rotation curves after subtracting the special relativistic boost which encodes the systemic motion of the AGN from the Earth. The maximum separation is less than $10$ km s$^{-1}$ and diminishes as $r$ increases.

We would like to point out that the present version of our general relativistic model can be applied as well to several megamaser accretion disks orbiting around SMBHs in a set of galaxies which has been studied by \citet{2013ApJ...767..154R} and \citet{2017ApJ...834...52G} in order to precisely estimate the central BH mass-to-distance ratio.

The estimation of the BH mass-to-distance ratio was made under the assumptions presented in Sec. II as a first calculation. A more detailed study would include taking into account the orbits of the systemic masers and their scatter about the LOS.

Further work is needed to suitably generalize our general relativistic method to include any kind of orbit geometry, in contrast to the circular orbits presented here, in order to implement it to the set of stars revolving around Sgr A* at the center of the Milky Way, for instance; we hope to push forward this issue in the near future.

In principle, it seems that this model is compatible with the existence of a disk bowl in the sector of systemic masers. Our formalism needs to be further developed in order to approach and analyze the effect of this bowl in the accretion disk, a sort of gravitational Lense-Thirring effect \citet{Bardeen-Peterson-Effect}, \citet{Lense-Thirring-Effect}. 


\section{Conclusions}

In this work we have derived from first principles a novel general relativistic model for studying Schwarzschild BH rotation curves that allows us to clearly distinguish and quantify the contribution of both the gravitational redshift and the special relativistic effects to the total redshift expression that can be linked to astrophysical observations. 

We applied this modeling, along with a Newtonian one, to fit the NGC 4258 observational data of 18 epochs spanning 3 years reported by \citet{2007ApJ...659.1040A}. Both the GR and Newtonian fits yield very similar, but excluding each other best estimation values for the parameters, however, GR is slightly preferred in the language of the $\chi^2_{\rm red}$, a tiny effect due to the fact that the difference between both models is less than $1\%$. Besides, in this situation one can rely on the fact that GR is the fundamental theory that better describes the classical gravitational interactions and phenomena, whereas Newtonian gravity is just a particular case of it.

This general relativistic modeling makes the analysis simpler since we are using a complete theory of gravity. We propose a new mass-to-distance ratio estimate for the SMBH in NGC 4258: $M/D = (0.5326\pm 0.00022)\times 10^7\; M_\odot{\rm Mpc^{-1}}$ and quantify the general and special relativistic effects for the closest to the SMBH masers. Moreover, we showed that the observed asymmetry in the BH rotation curves traced by megamasers on the accretion disk of this AGN is compatible with the magnitude of the gravitational redshift effect. However, the magnitude of this general relativistic effect is obfuscated by the errors in the red- and blueshift observations, preventing its unambiguous detection. Further precision in the positions of water masers is needed in order to overcome this hindrance towards the detection of this important general relativistic effect.


\acknowledgments
We acknowledge \citet{2007ApJ...659.1040A} for making  their NGC 4258 data compilation readily available for everyone and thank an anonymous referee for very enlightening and constructive reports on our manuscript.
A.H.-A., U.N. and O.L.-C. thank Sistema Nacional de Investigadores (SNI) for support. U.N. acknowledges support from CIC-UMSNH. 
The authors are grateful to FORDECYT-PRONACES-CONACYT for support of the present research under Grants No. CF-MG-2558591 and No. CF-140630. R. L.-C. also acknowledges support through a PhD grant No. 15706.

%

\vspace{5mm}
\facilities{VLBA}

\appendix

\section{Total shifts}\label{sec:appendix}

Here we compute the total shifts
\begin{equation}\label{ZT}
  1 + Z_{\rm tot_{1,2}} \equiv \frac{\omega_{e}^{\rm (Schw)}}{\omega_o^{\rm (boost)}} =
  \frac{\omega_e^{\rm (Schw)}}{\omega_o^{\rm (Schw)}}\frac{\omega_{o}^{\rm (Schw)}}{\omega_o^{\rm (boost)}} =
  \frac{\omega_e^{\rm (Schw)}}{\omega_o^{\rm (Schw)}}\frac{\omega_{e}^{\rm (boost)}}{\omega_o^{\rm (boost)}},
\end{equation}
where in the last equality we used the expression
\begin{equation}\label{Zequal}
  \omega_o^{\rm (Schw)} = \omega_e^{\rm (boost)};
\end{equation}
now, by using the definitions
\begin{eqnarray}\label{Zschw}
  1 + Z_{\rm Schw} &\equiv& \frac{\omega_e^{\rm (Schw)}}{\omega_o^{\rm (Schw)}} \\
  \label{Zboost}
  1 + Z_{\rm boost}&\equiv& \frac{\omega_{e}^{\rm (boost)}}{\omega_o^{\rm (boost)}}
\end{eqnarray}
in (\ref{ZT}) we obtain
\begin{equation}\label{ZT1}
  1 + Z_{\rm tot_{1,2}} = (1 + Z_{\rm Schw}) (1 + Z_{\rm boost}),
\end{equation}
where, by using the systemic velocity $v_0$ of NGC $4258$ for the boost parameters $\beta \equiv v_{0}/c$ and $\gamma \equiv (1-\beta^2)^{-1/2}$, we have for the shifts due to the boost
\begin{eqnarray}
  \label{Zboost2}
  1 + Z_{\rm boost}&=& \gamma (1 + \beta), 
\end{eqnarray}
while for the shifts of photons emitted by massive particles in circular orbits in the Schwarzschild spacetime we have
\begin{eqnarray}
  \label{Zschw1}
  1 + Z_{\rm Schw} &=& 1 + Z_{\rm grav} \pm Z_{\rm kin} 
\end{eqnarray}
with the expressions for the gravitational redshift given by (\ref{Zc}) and the kinematic blue/redshifts by 
\begin{eqnarray}
  \label{Zkin}
  Z_{\rm kin} = \epsilon \sin\theta_0 \sqrt{\frac{M r_e}{(r_e - 2M)(r_e - 3M)}}\,,
\end{eqnarray}
where we have taken into account the maser scatter $\epsilon$ along the azimuthal angle and the inclination of the disk in the polar angle of our modeling.

\subsection{Errors for the total shifts}

From Eq. (\ref{Zc}) we have the expression
\begin{eqnarray}
  \label{Zschw2}
  (1 + Z_{\rm grav})^{2} &=& \frac{r_e}{r_e - 3M}\,. 
\end{eqnarray}
Now, using (\ref{Zschw2}), we calculate the error associated to a measurement of the gravitational redshift
\begin{eqnarray}
  \label{errorGrav}
  \delta Z_{\rm grav} &=& - \frac{3}{2}(1 + Z_{\rm grav})^{3} \left(\frac{M}{r_e}\right) \left(\frac{\delta r_e}{r_e}\right), 
\end{eqnarray}
while using the expression (\ref{Zkin}) we obtain its associated error
\begin{eqnarray}
  \label{errorKin}
  \delta Z_{\rm kin} &=& \frac{1}{2} \epsilon \sin\theta_0 Z_{\rm kin}^{3} \left(\frac{6M^{2}-r_{e}^{2}}{Mr_{e}}\right) \left(\frac{\delta r_e}{r_e}\right);
\end{eqnarray}
by combining expressions (\ref{errorGrav}) and (\ref{errorKin}) we obtain the total error associated with 
$Z_{\rm tot_{1,2}}$: 
\begin{eqnarray}
  \label{errorTotal}
\delta Z_{\rm tot_{1,2}} &=& \left( \delta Z_{\rm grav} \pm \delta Z_{\rm kin} \right) (1 + Z_{\rm boost}).
\end{eqnarray}



\bibliography{NGC4258_ApJ.bib}{}
\bibliographystyle{aasjournal}




\end{document}